\documentclass[11pt]{article}

\usepackage[cp1251]{inputenc}

\usepackage{epsfig}

\newcommand{\N}{N\raise.7ex\hbox{\underline{$\circ $}}$\;$}

\begin{document}

\title{ E.M. Ovsiyuk,  N.V. Gutsko,  V.M.  Red'kov\footnote{e.ovsiyuk@mail.ru; gutsko-nv@yandex.ru; redkov@dragon.bas-net.by}\\[3mm]
     Transitivity in polarization optics \\and the diagonalization of quadratic forms\\[3mm]
     {\small Mozyr State Pedagogical University named after I.P. Shamyakin, Belarus\\
     B.I. Stepanov Institute of Physics, NAS of Belarus}
}

\maketitle

\begin{abstract}

Any one  measurement with  polarized
light makes it possible to fix the Mueller matrices of the Lorentz
type with up to four arbitrary numeric parameters $ (x, u; z, w) $. These
parameters are subject to the  quadratic condition.
 It is demonstrated that
the  quadratic form can be diagonalized;  in the case of partially polarized light
 four diagonal coefficients turn out to benon-zero and  positive; in the case of completely polarized
light two diagonal coefficients equal to zero.

 \end{abstract}

In [1, 2] it was shown that any one  measurement with  polarized
light makes it possible to fix the Mueller matrices of the Lorentz
type with up to four arbitrary numeric parameters $ x, u; z, w $. These
parameters are subject to the  quadratic condition
$$
 x^{2}   (A^{2}   - {\bf B}^{2} )
       + 2x y  A (B^{2}  -{\bf A}^{2} )
+
  y^{2}  [  (    {\bf A}^{2} +  {\bf B}^{2} - B^{2} ){\bf A}^{2}  - A^{2} B^{2}  ]
=
$$
$$
=    z^{2}  ( B^{2}   -  {\bf A}^{2} )        + 2 z w
  B (   A^{2} - {\bf B}^{2}    )   +
  w^{2}  [
  (  {\bf A}^{2} +  {\bf B}^{2}  - A^{2}  ) {\bf B}^{2}  -  A^{2} B^{2}  ] + 1 \;.
$$
$$
\eqno(1)
$$

\noindent The quantities $ A, {\bf A}, B, {\bf B} $ are defined by
the initial and final Stokes 4-vectors
$$
A = S+S', \qquad B = S -S', \qquad {\bf A} = {\bf S} + {\bf S}' ,
\qquad {\bf B} = {\bf S} - {\bf S}'
$$
$$
S> 0 \; , \qquad S' > 0 \; , \qquad S^{2} -{\bf S}^{2} =  S^{'2}
-{\bf S}^{'2}  = \mbox{inv} = + V^{2}  \geq 0
$$
$$
A^{2} = S_{0}^{2} + S_{0}^{'2} + 2 S_{0} S_{0}' = s^{2} + s^{'2}
+2s s' \; , \qquad
$$
$$
B^{2} = S_{0}^{2} + S_{0}^{'2} - 2 S_{0} S_{0}' = s^{2} + s^{'2}
-2s s' \; , \qquad
$$
$$
{\bf A}^{2} =
s^{2} + s^{'2}  - 2 V^{2} + 2 \sqrt{s^{2} - V^{2}} \sqrt{s^{'2} -
V^{2}} \cos \phi
$$
$$
{\bf B}^{2} =
s^{2} + s^{'2}  - 2 V^{2} - 2\sqrt{s^{2} - V^{2}} \sqrt{s^{'2} -
V^{2}} \cos \phi\;
\eqno(2)
$$

\noindent numeric parameters $ x, u; z, w $ determine  the parameters of possible Mueller matrices
of the Lorentzian type according to
$$
  n_{0}  = A x  -   {\bf A} ^{2}    y  \; , \qquad
  {\bf n} = z   {\bf A}   -   w  A  {\bf B}  + y {\bf A} \times {\bf B} \;  ,
$$
$$
m_{0}  = - B  z  +   {\bf B}^{2}  w   \; , \qquad {\bf m} =  x
{\bf B}    - y  B  {\bf A}   + w {\bf A} \times {\bf B}   \; .
\eqno(3)
$$

Note the different dimensions of the coordinates $x,y,z,w$:
$$
[x] =[S]^{-1} , \qquad  [y] =[S]^{-2} , \qquad  [z] =[S]^{-1} ,
\qquad  [w] =[S]^{-2} .
$$

\noindent
 All 6  terms in (1) have  such a structure
 such the dimensions of parameters $ x, z $ and $ y, w $ are mutually compensated
 with corresponding dimensions $ [S] $ and $ [S] ^ {2} $ variables formed from
 the initial and final Stokes 4-vectors, and therefore all the quantities in equation
 (1) are  dimensionless.

The task of the present paper is to explore the  geometry  of the surface (1)
in 4-dimensional space $(x,y,z,w)$.
In equation (1) we can see two quadratic forms: in variables   $(x,y)$ and  $(z,W)$,
so one cane transform them
  to a diagonal form:
$$
\left | \begin{array}{c} x
\\
y \end{array} \right | = \left | \begin{array}{cc}
\cos \alpha & \sin \alpha \\
- \sin \alpha  & \cos \alpha
\end{array} \right |
\left | \begin{array}{c} X \\
Y \end{array} \right |, \qquad \left | \begin{array}{c} z
\\
w \end{array} \right | = \left | \begin{array}{cc}
\cos \alpha' & \sin \alpha'  \\
- \sin \alpha' & \cos \alpha'
\end{array} \right |
\left | \begin{array}{c}
Z \\
W \end{array} \right |. \eqno(4)
$$

\noindent Substituting (4) into  (1)  and imposing  requirements for the diagonalization
of two quadratic forms, we  arrive at the following relations
$$
\mbox{tg}\; 2\alpha  = { 2 \; a  \over        c - b   }\; , \qquad
\mbox{tg}\; 2\alpha'  = { 2 \; a'  \over
      c'     - b'   }\; .
 \eqno(5)
$$
$$
X^{2} (  b \cos^{2} \alpha   - a \sin 2 \alpha + c \sin^{2} \alpha
) + Y^{2} (  b \sin^{2} \alpha   + a \sin 2 \alpha + c \cos^{2}
\alpha  )
  =
$$
$$
=Z^{2} (  b' \cos^{2} \alpha'    - a' \sin 2 \alpha' + c' \sin^{2}
\alpha'  )  + W^{2} (   b' \sin^{2} \alpha'    + a' \sin 2 \alpha' +
c ' \cos^{2} \alpha'   )   +1 \; ,
\eqno(6)
$$

\noindent
where the notation is used
$$
(A^{2}   - {\bf B}^{2} )=b \;, \qquad A (B^{2}  -{\bf A}^{2} )=a
\; , \qquad (    {\bf A}^{2} +  {\bf B}^{2} - B^{2} ){\bf A}^{2}
- A^{2} B^{2} =c \; ,
$$
$$
(B^{2}   - {\bf A}^{2} )=b'\;, \qquad B (A^{2}  -{\bf B}^{2} )=a'
\; , \qquad (    {\bf B}^{2} +  {\bf A}^{2} - A^{2} ){\bf B}^{2}
- A^{2} B^{2} =c' \; .
\eqno(7)
$$

Further, taking into account the elementary relations
$$
\cos 2\alpha = \pm {1 \over  \sqrt{1 + \mbox{tg}^{2} 2\alpha  } }
= \pm { ( c - b)  \over \sqrt{ ( c - b)^{2} +   4  a^{2} } } \;  ,
$$
$$
 \sin 2 \alpha =  \ \pm { 2a
\over \sqrt{ ( c - b)^{2} +   4  a^{2} } }\;,
$$
$$
\cos ^{2} \alpha ={ \sqrt{ ( c -
b)^{2} +   4  a^{2} }  \pm ( c - b)
 \over 2 \sqrt{ ( c - b)^{2} +   4  a^{2} } }  \; ,
 $$
 $$
\sin ^{2} \alpha = { \sqrt{ ( c -
b)^{2} +   4  a^{2} }  \mp ( c - b)
 \over 2 \sqrt{ ( c - b)^{2} +   4  a^{2} } }
$$

\noindent we arrive at (let $\delta, \delta ' = \pm 1$)
$$
X^{2} \left ( {b+c \over 2} - \delta  {\sqrt{(b+c)^{2} - 4(bc
-a^{2}) } \over 2} \right ) +
$$
$$
+
 Y^{2} \left ( {b+c \over 2} + \delta  {\sqrt{(b+c)^{2} -4(bc -
a^{2}) } \over 2} \right ) -
$$
$$
-Z^{2} \left ( {b'+c' \over 2} -  \delta '{\sqrt{(b'+c')^{2} -
4(b'c' -a^{'2}) } \over 2} \right ) -
$$
$$
-
 W^{2} \left ( {b'+c' \over 2} + \delta ' {\sqrt{(b'+c')^{2}
-4(b'c' - a^{'2}) } \over 2} \right ) =  1 \; . \eqno(8)
$$

First, we assume isotropic  Stokes vectors: $ V ^ {2} = 0 $. This will simplify the analysis, since
$$
A^{2} =  s^{2} + s^{'2} +2s s' \; , \qquad B^{2} =  s^{2} + s^{'2}
-2s s' \; , \qquad
$$
$$
{\bf A}^{2} = s^{2} + s^{'2}   + 2 s s'  \cos \phi \; , \qquad
{\bf B}^{2} =  s^{2} + s^{'2}   - 2 s s' \cos \phi \; .
$$

\noindent We calculate
$$
b= 2s s'(1  +  \cos \phi) \; , \qquad c = (s+s')^{2}   2 s s'  (1
+ \cos \phi ) \; .
$$
$$
b +c =   2s s'(1  +  \cos \phi)  +
 (s+s')^{2}   2 s s'  (1 + \cos \phi ) \geq 0 \; .
$$
$$
a^{2} =  (s^{2} + s^{'2} +2s s')  (2s s')^{2}  \;  (  1 + \cos \phi
)^{2}\; , \qquad
bc -a^{2} = 0 \; . \eqno(9a)
$$

\noindent Therefore, if  $ \delta = +1 $, in (8) the coefficient at
$ X ^ {2} $ is zero and the coefficient at $ Y ^ {2} $ is positive  $ (b + c)> 0 $.
When $ \delta = +1 $, the coefficient at $ Y ^ {2} $ is zero, and
the coefficient at $ X ^ {2} $ is positive  $ (b + c)> 0 $.

In this case (for definiteness, let $ \delta = +1 $)
$$
\cos^{2} \alpha = {c \over c+b }= {(s+s')^{2} \over 1 + (s+s')^{2}
}\;,
$$
$$
 \sin^{2} \alpha = {b \over c+b } ={1  \over 1 +
(s+s')^{2} }\; . \eqno(9b)
$$

Analyzing the coefficients of the $ Z ^ {2}, W ^ {2} $, we get
$$
b'= -2s s'(1  +  \cos \phi) < 0 \; , \qquad c'= - (s - s') ^{2}  2
s s'  (1 + \cos \phi )< 0  \; ,
$$
$$
b'  +c' =  -2s s'(1  +  \cos \phi)  -
 (s -  s')^{2}   2 s s'  (1 + \cos \phi ) \leq 0 \; .
$$
$$
a^{'2} =  (s-s')^{2}   (2s s')^{2}  \;  (  1 + \cos \phi )^{2}\; , \qquad
b'c' -a^{'2} = 0\;. \eqno(10a)
$$

\noindent Thus, in (8) the coefficient at $ Z ^ {2} $ is
zero and the coefficient at $ W ^ {2} $ is  negative $ b '+ c' \leq
0 $ (we assume  $ \delta '= +1 $). In this case
$$
\cos^{2} \alpha' = {c' \over c'+b' } ={ (s-s')^{2} \over 1 +
(s-s')^{2} } \;,
$$
$$
\sin^{2} \alpha'  = {b' \over c'+b' } = { 1
\over 1 + (s-s')^{2} }\; . \eqno(10b)
$$

Therefore, the set of Mueller matrices, connecting  two states
completely polarized light, are described by the relation
$$
0 \; X^{2} + (b+c) Y^{2}  - 0\; Z^{2} - (b'+c') W^{2} =  +1 \; .
\eqno(11a)
$$

The whole set of Mueller matrices, defined
by the  pair of isotropic Stokes vectors can be specified  by
following three independent variables:
$$
X,\; Z  \;\; \mbox{are arbitrary}, \qquad
 Y = { \cos \Gamma \over  \sqrt{ +(b+c)} } \; ,
 $$
 $$
 W = { \sin \Gamma \over  \sqrt{-( b'+c')} } \;, \qquad  \Gamma \in [0, 2 \pi ] \;  .
\eqno(11b)
$$

Let us consider another simple case. Let an initial beam
is the natural light, then
$$
S_{0} = s\; , \qquad {\bf S}=0\;  , \qquad V^{2} = s^{2} , \qquad
s' > s \; ,
$$
$$
A^{2} =  s^{2} + s^{'2} +2s s' \; , \qquad B^{2} =  s^{2} + s^{'2}
-2s s' \; ,
$$
$$
{\bf A}^{2} =  s^{'2}  -  s^{2} \; , \qquad {\bf B}^{2} =  s^{'2}
-  s^{2} \; .
 \eqno(12a)
 $$

\noindent We calculate
$$
b= 2s^{2}  +2s s' = 2s (s'+s) > 0 \; , \qquad
c =  2s (s' +s ) (   s'  -s   )^{2} > 0 \; ,
$$
$$
a^{2}  =  4s^{2} (s' +s)^{2} (s'-s)^{2} \; ,  \qquad
bc -a^{2} = 0 \; . \eqno(12b)
$$

\noindent The coefficient at $ X ^ {2} $ is zero, and that at
$ Y ^ {2} $ equals $ (b + c) $:
$$
b+c =  +  2s (s' +s )  [ 1 + (   s'  - s   )^{2}  ] \; ,
\eqno(12c)
$$

\noindent
in this case
$$
\cos^{2} \alpha = {c \over c+b }= {(s'-s)^{2} \over 1 + (s'-s)^{2}
}\;,
$$
$$
\sin^{2} \alpha = {b \over c+b } ={1  \over 1 +
(s'-s)^{2} }\; . \eqno(12d)
$$

Similarly, examine the coefficients at $ Z ^ {2}$ and $ W ^ {2}$:
$$
b'=  - 2s (s' -s) < 0
\; , \qquad
c' =  -2s (s + s' ) (   s^{'2}   - s ^{2}  ) < 0 \; ,
$$
$$
a^{'2}  =  4s^{2} (s' -s)^{2} (s'+s)^{2}, \qquad
b'c' -a^{'2} =  0\,. \eqno(13a)
$$

\noindent The coefficient at $ Z ^ {2} $ is equal to zero,
ant the coefficient at  $ W ^ {2} $ equals $ (b '+ c')$:
$$
b'+c' =  -  2s (s' -s )  [ 1 + (   s'  +s   )^{2}  ] \,,
\eqno(13b)
$$

\noindent in this case
$$
\cos^{2} \alpha' = {c' \over c'+b' }= {(s'+s)^{2} \over 1 +
(s'+s)^{2} }\;,
$$
$$
 \sin^{2} \alpha'  = {b' \over c'+b'  } ={1
\over 1 + (s'+s)^{2} }\; . \eqno(13c)
$$

The basic quadratic equation takes the form
$$
0 \; X^{2} + (b+c) Y^{2}  - 0\; Z^{2} - (b'+c') W^{2} =  +1 \; ,
\eqno(14a)
$$
$$
+(b+c ) =    2s (s' +s )  [ 1 + (   s'  -s   )^{2}  ] \; ,
$$
$$
-(b'+c') =    2s (s' -s )  [ 1 + (   s'  +s   )^{2}  ]\; .
\eqno(14b)
$$

Let us consider another simple case,  when an initial beam is
partially polarized, and the final beam is the natural
light:
$$
s'= V\; , \qquad s > s' \;, \qquad
{\bf A}^{2} = s^{2} - s^{'2} \; , \qquad {\bf B}^{2} =  s^{2} -
s^{'2}\; ,
$$
$$
A^{2} =  s^{2} + s^{'2} +2s s' \; , \qquad B^{2} =  s^{2} + s^{'2}
-2s s' \; .
$$
$$
b = + 2s'(s +s')  > 0 \;, \qquad
b'= 2s^{'2} -2s s' = - 2s' (s-s') < 0 \; ,
$$
$$
c = 2s' (s-s') (s^{2} - s^{'2} )> 0\;, \qquad
c' =  -2s' (s+s') (s^{2} - s^{'2} )< 0 \; ,
$$
$$
a^{2} = 4s^{'2} (s+s')^{2} (s'
- s  )^{2}  \; , \qquad
a^{'2} =  4s^{'2} (s-s')^{2} (s' + s )^{2}\; .
\eqno(15a)
$$

We see that the equalities
$bc -a^{2} = 0 \;, \;b'c' - a^{'2} =0$.
The basic quadratic equation takes the form
$$
0 \; X^{2} + (b+c) Y^{2}  - 0\; Z^{2} - (b'+c') W^{2} =  +1 \; .
\eqno(15b)
$$

Let us consider  another simple case: a set of Mueller matrices that do not change of light intensity
$
s' =s$;
in this case
$$
A^{2} =  4s^{2}\; , \qquad B^{2} =  0 \; ,
$$
$$
{\bf A}^{2} = 2(s^{2}  -  V^{2})(1  +
 \cos \phi )\; ,
 $$
 $$
{\bf B}^{2} =  2 (s^{2}  -  V^{2})(1 -  \cos \phi ) \; .
\eqno(16a)
$$

The basic quadratic equation takes the form
$$
 x^{2}   (A^{2}   - {\bf B}^{2} )
       - 2x y  A {\bf A}^{2}
+
  y^{2}    (    {\bf A}^{2} +  {\bf B}^{2}  ){\bf A}^{2}
+
$$
$$
+ 2 (s^{2} -V^{2}) (1 + \cos \phi ) \;  Z^{2}
 +
 $$
 $$
 +
 8V^{2} (s^{2} - V^{2})  (1 - \cos \phi)  \;  W^{2}
 =  + 1  \; .
\eqno(16b)
$$

We need only to diagonalize the first quadratic form. After calculation we get
$$
b = 2 \left [ V^{2} ( 1 - \cos \phi) + s^{2} (1+ \cos \phi )  \right ]  > 0 \; ,
$$
$$
c =
  8 ( s^{2} - V^{2})^{2}  ( 1 +  \cos \phi  ) > 0 \; ,
$$
$$
a^{2} = 16s^{2} (s^{2}-V^{2})^{2} ( 1 + \cos \phi)^{2}\, ,
$$
$$
bc -a^{2} = 16\,V^{2}\,\sin^{2}\phi\, (s^{2}-V^{2})^{2}  > 0 \; .
\eqno(16c)
$$

It follows that the first quadratic form is also reduced to diagonal form with positive coefficients.

A situation where the initial and final light beams
 are partially polarized, is the most difficult.
  The coefficients of a diagonalized quadratic form
  can be calculated explicitly and  they are positive.

Let us detail this subject as well. First, We find
$$
b=   2V^{2}+ 2s s' \left  (1 +  \sqrt{1 - V^{2}/s^{2} } \sqrt{1 - V^{2}/ s^{'2}} \cos \phi \right ) >
2V^{2} \geq 0 \; ,
$$
$$
c =
 [ (s+s')^{2}-4V^{2} ]\, \times
 $$
 $$
 \times \left [ (s-s')^{2} -2 V^{2}+ 2 ss' \left  (
 1 + \sqrt{1 - V^{2}/ s^{2} } \sqrt{1 - V^{2}/s^{'2} }  \cos \phi \right ) \right ] -
 $$
 $$
 -
 (s+s')^{2} (s-s')^{2} \; .
$$

In turn, we obtain
$$
b+c = \left (2ss' + 2  \cos \phi \sqrt{s^{2}-V^{2}}  \sqrt{s^{'2}-V^{2}} \right )
   \left   ( 1 +  [ (s+s')^{2}- 4V^{2}]  \right )  +
     $$
     $$
     + 2V^{2} - 2 V^{2} [ (s+s')^{2}- 4V^{2}]   -4V^{2}  (s-s')^{2}\;.
\eqno(17a)
$$

At  $\phi =\pi$ we have the minimal values for $(b+c)$:
$$
(b+c)_{min}  = \left (2ss' - 2  \sqrt{s^{2}-V^{2}}  \sqrt{s^{'2}-V^{2}} \right )
   \left   ( 1 +  [ (s+s')^{2}- 4V^{2}]  \right )  +
     $$
     $$
     + 2V^{2} \left   ( 1 -  [ (s+s')^{2}- 4V^{2}]  \right )
        -4V^{2}  (s-s')^{2}\; ;
\eqno(17b)
$$

\noindent
Numerical calculation shows that the quantity  $(b+c)_{min}$ is positive  (see Fig.1)

\begin{figure*}[hbt]
\hspace{30mm}
\includegraphics[width=10cm]{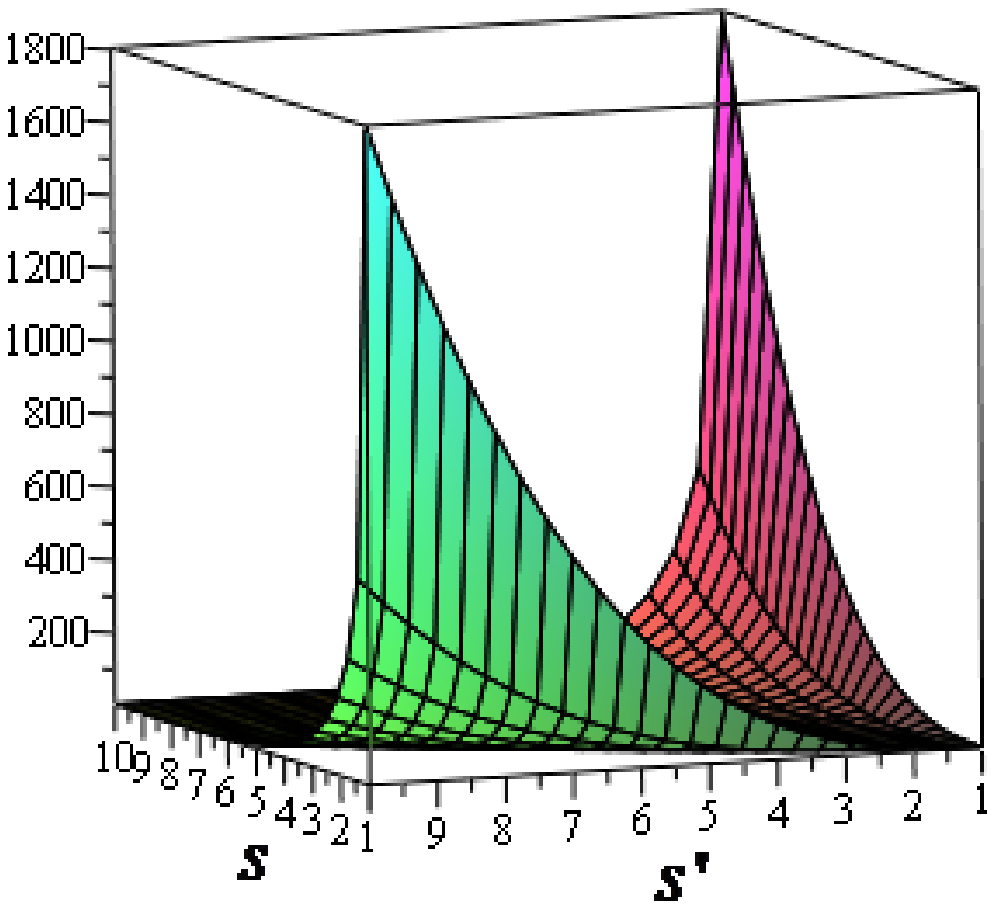}

\end{figure*}

\vspace{-30mm}

\begin{center}
Fig. 1: $(b+c)_{min} (V=1)$
\end{center}

Additionally let us calculate
$$
a^{2} = (s^{2} + s^{'2}      +2s s')\,(2V^{2}-2s s'-2 \sqrt{s^{2} - V^{2}} \sqrt{s^{'2} - V^{2}} \cos \phi)^{2}\,.
\eqno(18a)
$$
and
$$
bc -a^{2} = -16\,V^{2}\,\sin^{2}\phi\,(s^{2}V^{2}+s^{'2}V^{2}-s^{2}s^{'2}-V^{4})=
$$
$$
=16\,V^{2}\,\sin^{2}\phi\, (s^{2}-V^{2})(s^{'2}- V^{2})  > 0
\eqno(18b)
$$

Turning to coefficients at  $X^{2}, Y^{2}$ in  (8)
$$
X^{2} \left ( {b+c \over 2} - \delta  {\sqrt{(b+c)^{2} - 4(bc -a^{2}) } \over 2} \right ) +
$$
$$
+
Y^{2} \left ( {b+c \over 2} + \delta  {\sqrt{(b+c)^{2} -4(bc - a^{2}) } \over 2} \right ) - ... = 1
\eqno(18c)
$$

\noindent
we conclude that they both are positive

Now we analyze coefficient at  $Z^{2}, W^{2}$. After simple calculation we get
$$
b'= -2s s'+2V^{2}-2\cos \phi  \sqrt{s^{2} - V^{2}} \sqrt{s^{'2} - V^{2}} =
$$
$$
c' = \left (s^{2}+s^{'2}-4V^{2}-2ss' \right )\,
\left (s^{2}+s^{'2}-2V^{2}-2 \cos \phi \sqrt{s^{2} - V^{2}} \sqrt{s^{'2} - V^{2}}  \right )-
$$
$$
- (s^{2}+s^{'2}+2s s')\,(s^{2}+s^{'2}-2s s') \; .
$$
$$
b'  +c'  =
 -\left (2ss' + 2  \cos \phi \sqrt{s^{2}-V^{2}}  \sqrt{s^{'2}-V^{2}} \right )
   \left   ( 1 +  [ (s-s')^{2}- 4V^{2}]  \right )  +
     $$
     $$
     + 2V^{2} - 2 V^{2} [ (s-s')^{2}- 4V^{2}]   -4V^{2}  (s+s')^{2}\;.
\eqno(19a)
$$

Numerical calculation shows that the quantity
$(b'+c')_{max}$  is negative (see Fig.2)

\begin{figure*}[hbt]
\hspace{+30mm}
\includegraphics[width=10cm]{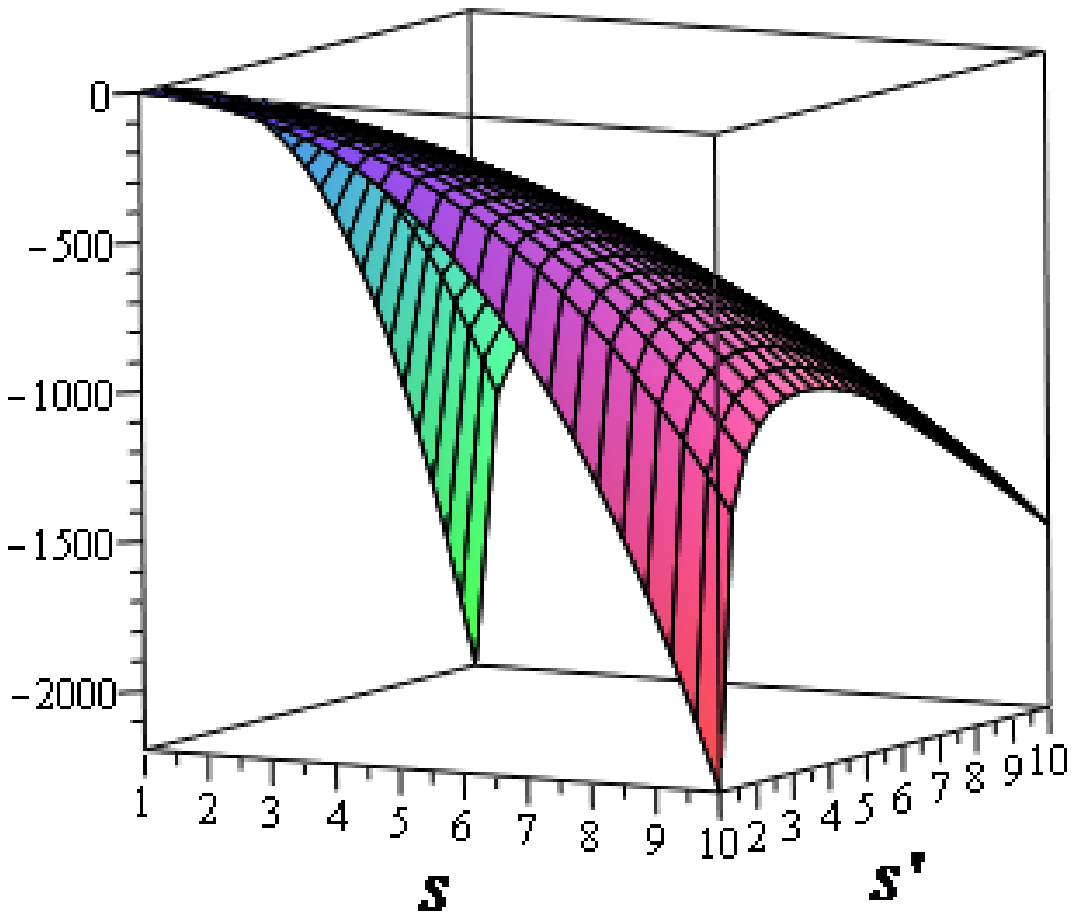}

\end{figure*}

\vspace{-30mm}
\begin{center}
Fig. 2: $(b'+c')_{max}(V=1)$

\end{center}

Additionally we find
$$
a^{'2} =  \left (s^{2} + s^{'2}      -2s s' \right )
\, \left (2V^{2}+2s s'+2 \cos \phi  \sqrt{s^{2} - V^{2}} \sqrt{s^{'2} - V^{2}}  \right )^{2}\,.
$$
$$
b'c' -a^{'2} = -16\,V^{2}\,\sin^{2}\phi\,(s^{2}V^{2}+s^{'2}V^{2}-s^{2}s^{'2}-V^{4}) =
$$
$$
=16\,V^{2}\,\sin^{2}\phi\, (s^{2}-V^{2})(s^{'2}- V^{2})  > 0
\eqno(19b)
$$

Therefore, coefficients at $Z^{2} $ $W^{2}$ in (8) are positive
$$
...
-Z^{2} \left ( {b'+c' \over 2} -  \delta '{\sqrt{(b'+c')^{2} - 4(b'c' -a^{'2}) } \over 2} \right ) -
$$
$$
-
W^{2} \left ( {b'+c' \over 2} + \delta ' {\sqrt{(b'+c')^{2} -4(b'c' - a^{'2}) } \over 2} \right ) =  1 \; .
\eqno(19c)
$$

{\bf Conclusion:}
Any one  measurement with  polarized
light makes it possible to fix the Mueller matrices of the Lorentz
type with up to four arbitrary numeric parameters $ (x, u); (z, w) $. These
parameters are subject to the  quadratic condition.
 It is demonstrated that
both quadratic forms, in variables $(x,y)$ and $(z,w)$, can be diagonalized;  in the case of partially polarized light
 four diagonal coefficients turn out to be positive; in the case of completely polarized
light two diagonal coefficient equal to zero.

\vspace{10mm}

%\newpage

{\bf REFERENCES}

\vspace{2mm}

1.
V.M. Red'kov, E.M. Ovsiuyk. Transitivity in the theory of
the Lorentz group and the Stokes - Mueller formalism in optics.
// Proceedings of 15th International School \& Conference "Foundation \& Advances in Nonlinear Science.
 Eds.: Kuvshinov V.I., Krylov G.G., September  19-22, 2010, Minsk. P. 1-27.
  - http://arxiv.org/abs/1005.4212. - [math-ph; physics.optics]; 22 pages.

2.
V.M. Red'kov, E.M. Ovsiuyk.
On determining Mueller matrices of an optical element by results of polarization measurements
// Proceedings of of III-th International Conference
"Optics of inhomogeneous structures-2011", 16-17 February 2011, Mogilev. Eds. B.A. Karpenko and others
P. 32--35.

\end{document}